# Magnetic Design of Superconducting Magnets

*E. Todesco[1],*
CERN, Geneva, Switzerland

**Abstract**
In this paper we discuss the main principles of magnetic design for superconducting magnets (dipoles and quadrupoles) for particle accelerators. We give approximated equations that govern the relation between the field/gradient, the current density, the type of superconductor (Nb–Ti or $Nb_3Sn$), the thickness of the coil, and the fraction of stabilizer. We also state the main principle controlling the field quality optimization, and discuss the role of iron. A few examples are given to show the application of the equations and their validity limits.

*Keywords*: magnets for accelerators, superconducting magnets, magnet design.

## 1  Introduction

The common thread of these notes is to provide some analytical guidelines with which to outline the design of a superconducting accelerator magnet. We consider this for both dipoles and quadrupoles: the aim is to understand the trade-offs between the main parameters such as the field/gradients, the free aperture, the type of superconductor, the operational temperature, and the current density. These guidelines are rarely treated in handbooks: here, we derive a set of equations that provides us with the main picture, with an error that can be as low as 5%. This initial guess can then be used as a starting point for fine tuning by means of a computer code to account for other field quality effects not discussed here, such as iron saturation, persistent currents, eddy currents, etc. Examples will be given in the text to guide the reader around the equations.

In Section 2, we introduce the sector coil design for electromagnets, derive the multipolar expansion, and write down the conditions for field quality that must be imposed on the wedges and on the angles of a sector coil. We also present the main equations for an electromagnet, relating the central field and the gradient to the current density and the coil width. In Section 3, we discuss the superconducting case; first we propose simple expressions for the critical surfaces of Nb–Ti and $Nb_3Sn$. We then focus on the peak field in a sector coil; this allows us to derive a complete equation giving the relation field/gradient versus coil width, aperture, superconductor type, and quantity of stabilizer. In Section 4, we analyse the role of iron, and discuss its beneficial effects on lowering the current density and keeping the magnetic flux within the magnet. In Section 5, we discuss an alternative design based on block coils, and in Section 6 we summarize the different steps taken in magnet design.

## 2  Electromagnets based on sector coils

In this section, we discuss the main equations for an electromagnet, starting with the Biot–Savart law, which yields the field produced by a current line. The section deals with a generic electromagnet, i.e. it is independent of the specific features of superconductivity: we only consider how to make a 'good' dipole or quadrupole field with current lines carrying a given current density. After having defined the multipolar expansion for a current line, we focus on the computation of the field given by a sector coil producing a dipolar field. This is the main building block that can be used to construct the so-called cos-theta layout, based on a keystoned cable and wedges. We then derive the conditions for the positions and number of wedges required to satisfy the accelerator field quality constraints. The quadrupole case is also considered, to show the similarities and the differences.

---
[1] ezio.todesco@cern.ch

## 2.1 Biot–Savart law and the multipole expansion

An infinitesimal current line d$r$ produces an infinitesimal magnetic field d$B$ according to the equation

$$\mathrm{d}\vec{B}(\vec{r}) = \frac{\mu_0}{4\pi} \frac{\vec{I} \times \mathrm{d}\vec{r}}{|r|^3}, \tag{1}$$

where $\vec{I}$ is vector of the current intensity, $r$ is the distance from the current line, and $\mu_0 = 4 \times 10^{-7}$ (N·A$^{-2}$) is the vacuum permeability. For a line in the $s$ direction of infinite length, integration yields

$$\vec{B}(\vec{r}) = \frac{\mu_0}{2\pi} \frac{\vec{I} \times \vec{r}}{|r|^2}, \tag{2}$$

and the problem becomes two-dimensional. The two components are given by

$$\begin{aligned}
B_x(x,y) &= -\frac{\mu_0 I}{2\pi} \frac{y - y_0}{(x - x_0)^2 + (y - y_0)^2}, \\
B_y(x,y) &= \frac{\mu_0 I}{2\pi} \frac{x - x_0}{(x - x_0)^2 + (y - y_0)^2},
\end{aligned} \tag{3}$$

and, using the complex notation $z = x + \mathrm{i}y$ and $B = B_y + \mathrm{i}B_x$ (see Fig. 1), we obtain

$$B_y(x,y) + \mathrm{i}B_x(x,y) = \frac{\mu_0 I}{2\pi} \frac{(x - x_0) - \mathrm{i}(y - y_0)}{(x - x_0)^2 + (y - y_0)^2}. \tag{4}$$

Since

$$\frac{a - \mathrm{i}b}{a^2 + b^2} = \frac{a - \mathrm{i}b}{(a + \mathrm{i}b)(a - \mathrm{i}b)} = \frac{1}{(a + \mathrm{i}b)}, \tag{5}$$

the field can be written as

$$\boxed{B(z) = \frac{\mu_0 I}{2\pi(z - z_0)}}, \tag{6}$$

where the current line is in $z_0$ and the field is given at position $z$.

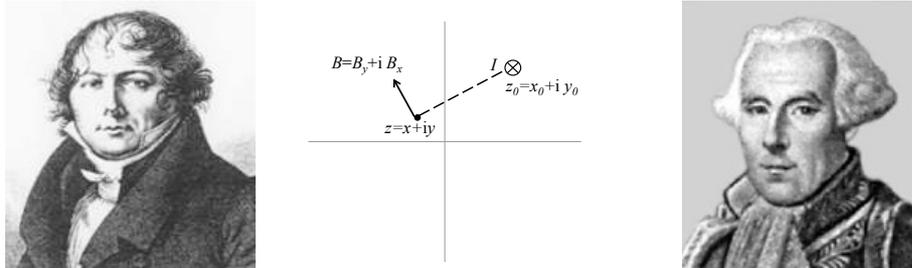

**Fig. 1:** Jean-Baptiste Biot (left); the field generated in $z$ by an infinite current line placed at $z_0$ (center); and Felix Savart (right).

The above formulation, based on complex notation, besides being compact, can easily produce a multipolar expansion: because

$$\frac{1}{1-z} = 1 + z + z^2 + z^3 + \ldots = \sum_{n=1}^{\infty} z^{n-1} \quad \text{for } |z| < 1 \tag{7}$$

(remember: the expansion is valid inside the unitary circle), one can write

$$B(z) = \frac{\mu_0 I}{2\pi(z-z_0)} = -\frac{\mu_0 I}{2\pi z_0} \frac{1}{1-\frac{z}{z_0}} = -\frac{\mu_0 I}{2\pi z_0} \sum_{n=1}^{\infty} \left(\frac{z}{z_0}\right)^{n-1} \quad \text{for } |z| < |z_0|. \tag{8}$$

When the multipolar expansion is carried out, we lose something: Eq. (6) is valid everywhere (except at the singularity in $z_0$), but the right-hand side of Eq. (8) is valid only for $|z|<|z_0|$). So this expression is good for estimating the field seen by the beams, but not for estimating the field in the magnet coil or in the iron.

In any current-free region $D$ the field has null divergence, and can be expanded in multipoles:

$$B(x,y) = \sum_{n=1}^{\infty} (B_n + iA_n)(x+iy)^{n-1} \quad x, y \in D. \tag{9}$$

To avoid coefficients with physical dimensions depending on the order of the multipoles, we usually define a reference radius $R_{\text{ref}}$. Moreover, the main component is factorized so that the normalized multipoles ($b_n$, $a_n$) become dimensionless (see, e.g., Ref. [1], p. 50); for a dipole,

$$\boxed{B(x,y) = 10^{-4} B_1 \sum_{n=1}^{\infty} (b_n + ia_n) \left(\frac{x+iy}{R_{\text{ref}}}\right)^{n-1}} \quad x, y \in D. \tag{10}$$

Here, $b_n$ and $a_n$ are called the normal and skew multipoles, respectively. The reference radius $R_{\text{ref}}$ has no physical meaning (it is just a choice of units to express the multipoles, and has nothing to do with the convergence radius of the series), and is usually chosen to be 2/3 of the aperture. For an accelerator magnet, typical acceptable field quality deviations from the ideal field are of the order of 0.01% at 2/3 of the aperture radius. This is why a $10^{-4}$ is also factorized in Eq. (10). In this way the normalized multipoles have values of the order of unity, providing a more compact expression and simplifying the compilation of tables (no exponential format is needed).

Comparing Eqs. (8) and (10), we obtain the main field of a current line,

$$B_1 = -\frac{I\mu_0}{2\pi} \operatorname{Re}\left(\frac{1}{z_0}\right) = -\frac{I\mu_0}{2\pi} \frac{x_0}{x_0^2 + y_0^2}, \tag{11}$$

and the multipoles are

$$\boxed{\begin{aligned} B_n + iA_n &= -\frac{I\mu_0}{2\pi z_0} \left(\frac{R_{\text{ref}}}{z_0}\right)^{n-1}, \\ b_n + ia_n &= -\frac{I\mu_0}{2\pi z_0} \frac{10^4}{B_1} \left(\frac{R_{\text{ref}}}{z_0}\right)^{n-1}. \end{aligned}} \tag{12}$$

This notation may seem a bit obscure, but in the following we will see how easily the conditions for field quality in a electromagnet can be derived from these expressions.

Equations (12) have a meaningful implication: since the position of the current line $z_0$ is larger than the reference radius $R_{\text{ref}}$, the argument of the power is smaller than one, and the multipoles vanish as a power law. Plotting the logarithm of the absolute value of the multipoles against the multipole order, one obtains a linear regression with slope $R_{\text{ref}}/|z_0|$ (see Fig. 2). The aim of magnetic design (see Refs. [1], pp. 45–63, [2]) is to arrange the current lines around the free aperture so that the contributions (12) compensate for each other. The target is to have multipoles of a few units for low order, $n \leq 3$, and less than one unit for the higher orders.

The requirements on field quality can be translated into the requirements on the positioning of the current lines in the cross-section; typically, the required precision is of the order of 0.1 mm. The multipole variation created by a misplacement of the current line of $\Delta z_0$ is given by

$$\Delta(B_n + iA_n) = n \frac{I\mu_0}{2\pi z_0^2} \left(\frac{R_{\text{ref}}}{z_0}\right)^{n-1} \Delta z_0 , \qquad (13)$$

and has the same power law decay as a current line. This implies that the random components of the field quality due to the position of the current lines also have a power law decay. This can be used to judge the precision of the measurement, corresponding to the appearance of a plateau where data no longer follow the Biot–Savart decay (see Fig. 2, right).

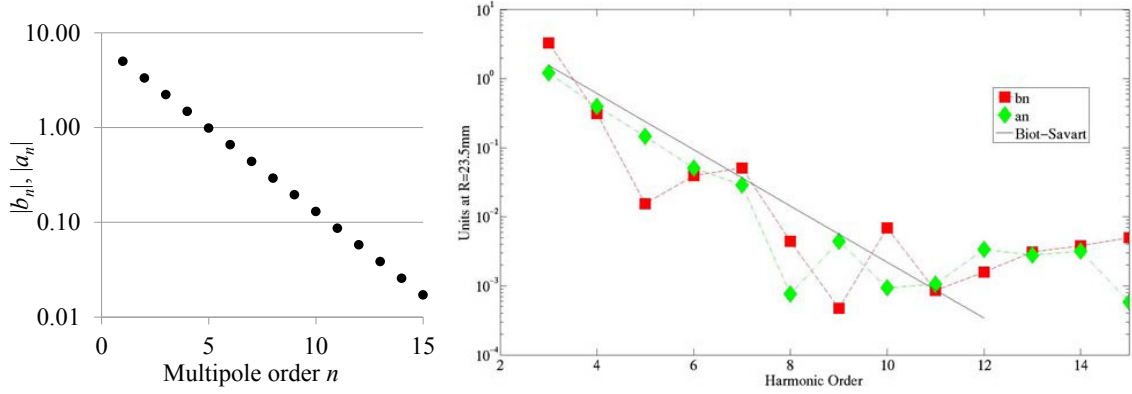

**Fig. 2:** Decay of multipoles of a current line with order plotted on a semilogarithmic scale (left), and an example of a magnetic measurement (right) of the random components of field quality in a Nb$_3$Sn short model HQ, showing the measurement precision at around $10^{-2}$ units [3].

## 2.2 Dipoles based on sector coils

It is well known that a cos-theta distribution of current density provides a perfect dipolar field (see, e.g. Refs., [1], pp. 45–63, [2, 4]). This means that there is a maximal current on the midplane, with opposite sign on the left and right sides of the aperture, and zero current at 90° and 270° (see Fig. 3, left). This can be approximated in a real winding by a sector coil with layers and wedges (see Fig. 4). Here we will present analytical expressions for the sector coil (see Fig. 3, right). Sector coils provide a very good approximation of an accelerator magnet, allowing a quasi-analytical approach that shows all the main features of magnet design [5, 6]. Moreover, equations can be easily implemented in spreadsheets, and solutions can be used as a starting point for more refined computations with finite element codes and/or optimizers.

A current line in a sector coil of width $w$ and angle $\alpha$, without wedges (see Fig. 3, right), yields a main field

$$B_1 = -\frac{I\mu_0}{2\pi}\text{Re}\left(\frac{1}{z_0}\right) = -\frac{I\mu_0}{2\pi}\frac{\cos\theta}{\rho} , \qquad (14)$$

where we have used polar coordinates $z = \rho \exp(i\theta)$. Replacing the current $I$ by the integration element, we find the equation for the main field as follows:

$$I \to j\rho \,d\rho \,d\theta \qquad \boxed{B_1 = -2\frac{j\mu_0}{2\pi}\int_{-\alpha}^{\alpha}\cos\theta \,d\theta \int_{r}^{r+w}\frac{\rho \,d\rho}{\rho} = -\frac{2j\mu_0}{\pi}w\sin\alpha} . \tag{15}$$

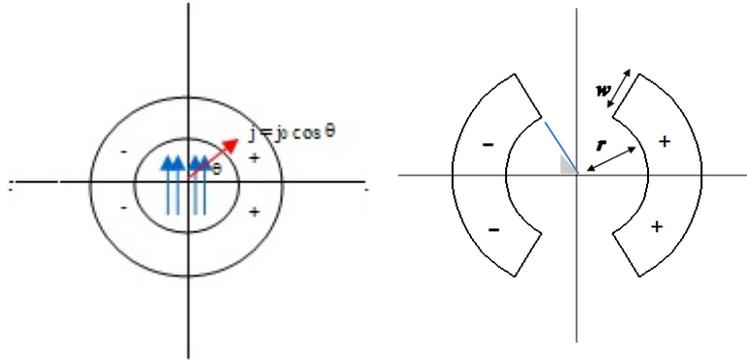

**Fig. 3:** A cos-theta distribution of current (left), and a sector coil without wedges (right)

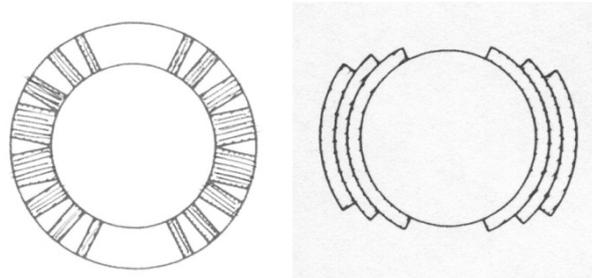

**Fig. 4:** Approximations of cos-theta coil with sector coils; wedges (left) and layers (right) [4]

So, for a dipole sector coil of aperture $r$ and coil width $w$, we have that

- the field is proportional to the current density $j$;
- the field is proportional to the coil width $w$;
- the field is independent of the coil aperture $r$.

We can verify that the configurations shown in Figs. 3 and 4 have the following expansion:

$$B = B_1\left(1 + 10^{-4}\left[b_3\frac{z^2}{R_{\text{ref}}^2} + b_5\frac{z^4}{R_{\text{ref}}^4} + b_7\frac{z^6}{R_{\text{ref}}^6} + \ldots\right]\right), \tag{16}$$

since the top–bottom symmetry guarantees $a_{2n+1} = 0$, the left–right antisymmetry guarantees that $b_{2n} = 0$, and the up–left/down–right antisymmetry guarantees that $a_{2n} = 0$. The computations are a little cumbersome, but are a good exercise in familiarizing oneself with the multipolar expansion. The odd normal components are called the 'allowed' multipoles in a dipole. We can see now the full force of the multipolar expansion: the conditions for a perfect dipole are reduced from a two-dimesional vectorial field to a few real coefficients ($b_3$, $b_5$, $b_7$,…). Usually a truncation at order 11 is used for tracking particles, so in the end the field quality condition for a dipole corresponds to cancelling only five coefficients.

For the same sector dipole as that just analysed, the integration of high-order multipoles $n > 2$ (the case $n = 2$ will be considered in the quadrupole subsection) yields

$$B_n = -\frac{j\mu_0 R_{\text{ref}}^{n-1}}{\pi} \frac{2\sin(n\alpha)}{n} \frac{(r+w)^{2-n} - r^{2-n}}{2-n} = -\frac{j\mu_0}{\pi} r \left(\frac{R_{\text{ref}}}{r}\right)^{n-1} \frac{2\sin(n\alpha)}{n(2-n)} \left[\frac{1}{\left(1+\frac{w}{r}\right)^{n-2}} - 1\right]. \quad (17)$$

Therefore, a sector coil of 60° has $b_3 = 0$, i.e. the first-order non-linearity is cancelled. The main field is given by

$$B_1 = -\frac{\sqrt{3} j\mu_0}{\pi} w \sim 6.9 \times 10^{-7} j\left[\text{A} \cdot \text{m}^{-2}\right] w[\text{m}] = 6.9 \times 10^{-4} j\left[\text{A} \cdot \text{mm}^{-2}\right] w[\text{mm}]. \quad (18)$$

The 60° sector coil is the equivalent of the Helmholtz coil, i.e. the simplest winding with a 'good' field quality. Unfortunately this is not 'good' enough since $b_5$ is too large. We have

$$b_5 = 10^4 \times \frac{1}{15} \frac{\sin 5\alpha}{\sin \alpha} \left(\frac{R_{\text{ref}}}{r}\right)^4 \frac{r}{w} \left[\frac{1}{\left(1+\frac{w}{r}\right)^3} - 1\right], \quad (19)$$

which for a typical case of $w = r$ and $\alpha = 60°$ gives $b_5 \sim 100$ units, whereas the requirements from particle dynamics are a few units at most. In the limit of a thin coil, $w \to 0$, we obtain the simplified expression

$$b_5 = 10^4 \times \frac{1}{5} \frac{\sin 5\alpha}{\sin \alpha} \left(\frac{R_{\text{ref}}}{r}\right)^4, \quad (20)$$

as given in Ref. [1], p. 53.

The interesting observation that can be derived from this simple case is that for $w/r \gg 1$ normalized multipoles are proportional to $r/w$. This means that a dipole with a thick coil $w$ w.r.t. to a small aperture $r$ will have 'naturally low' multipoles. Conversely, a dipole with large aperture and low field (as in the Relativitic Heavy Ion Collider main dipoles [7], where $r/w \sim 0.25$) will have a field quality that is more difficult to optimize.

To set two multipoles to zero requires two free parameters. The designers of the Tevatron [8] chose to have a two-layer sector coil with the same cable width; see Fig. 5. Therefore, $B_3$ and $B_5$ could be set to zero:

$$\begin{aligned} B_3 &\propto \sin(3\alpha_1)\left[\frac{1}{r+w} - \frac{1}{r}\right] + \sin(3\alpha_2)\left[\frac{1}{r+2w} - \frac{1}{r+w}\right] = 0, \\ B_5 &= \sin(5\alpha_1)\left[\frac{1}{(r+w)^3} - \frac{1}{r^3}\right] + \sin(5\alpha_2)\left[\frac{1}{(r+2w)^3} - \frac{1}{(r+w)^3}\right] = 0. \end{aligned} \quad (21)$$

This is a system of non-linear equations in the angles $\alpha_1$ and $\alpha_2$. It has solutions only for $w/r < 0.5$ (see Fig. 6); this is indeed the case for the Tevatron main diplole coil, which has $w/r \sim 0.2$, corresponding to the angles 37° and 72°. Using the above expressions, one can compute that this produces $b_7 = 20$ units, which in Tevatron times was considered not to be critical. During the 1980s, the concern for high-order non-linearities became more and more important. SSC [9], HERA [10], and LHC [11] dipole coils make use of wedges, leading to more free parameters, better to optimize the field quality (see Fig. 7, left).

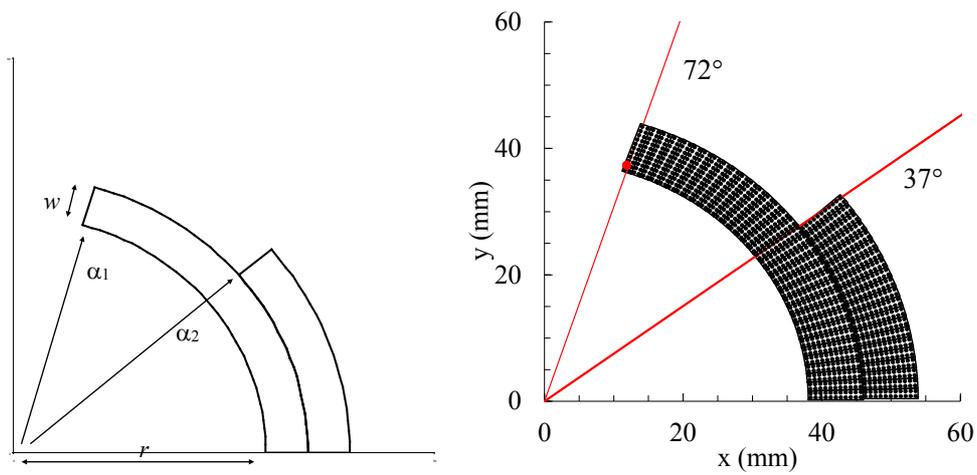

**Fig. 5:** Coil layout with two layers and no copper wedges (left), and coil used for Tevatron main dipoles (right)

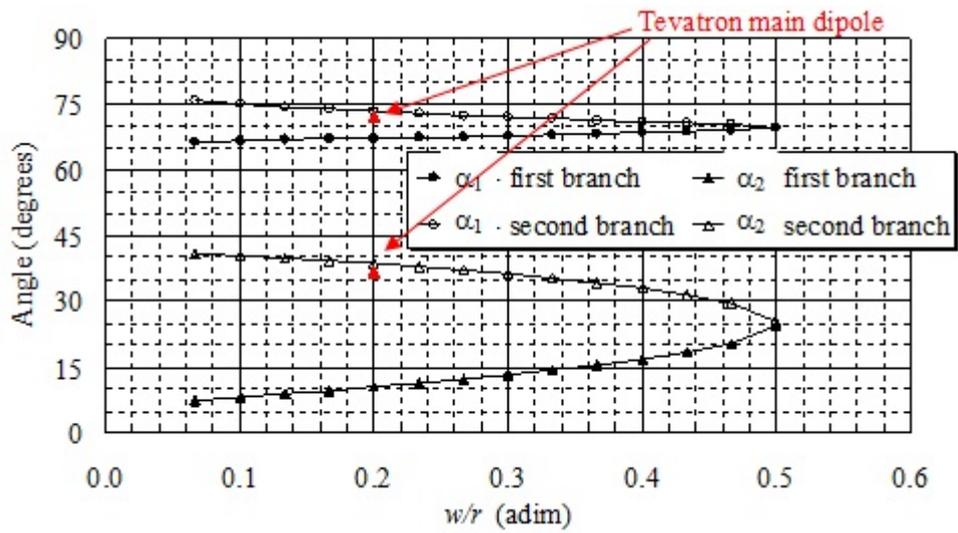

**Fig. 6:** The solutions for the angles of the inner layer $\alpha_1$ and the outer layer $\alpha_2$ vs. the ratio $w/r$ and the coil adopted in Tevatron dipoles.

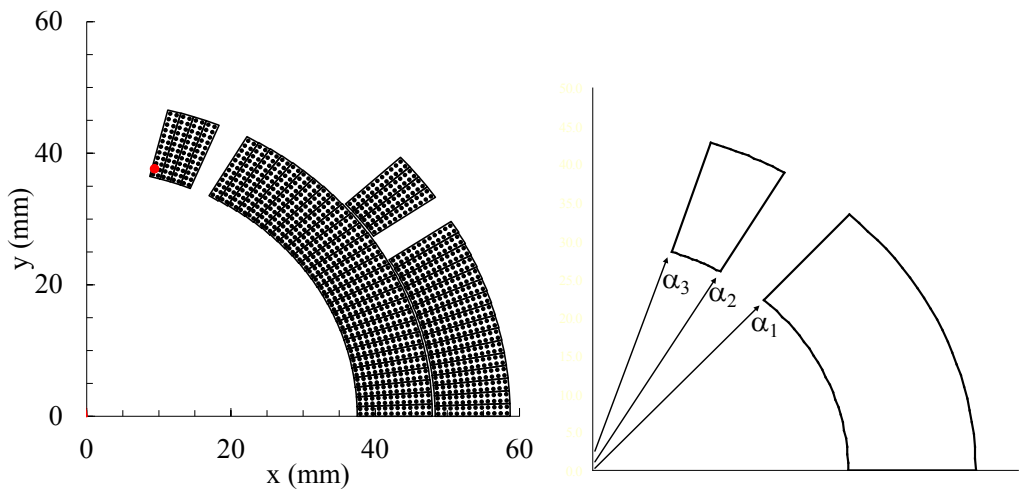

**Fig. 7:** Coil of the main HERA dipole (left) and coil layout with one layer and one wedge (right)

With one layer and one wedge (see Fig. 7, right), there are three free parameters $\alpha_1$, $\alpha_2$, and $\alpha_3$, and in the non-normalized multipole equations we simply have to add the expression for each sector coil:

$$B_3 = \frac{\mu_0 j R_{\text{ref}}^2}{\pi} \frac{\sin(3\alpha_1) - \sin(3\alpha_2) + \sin(3\alpha_3)}{3} \left[\frac{1}{r+w} - \frac{1}{r}\right],$$

$$B_5 = \frac{\mu_0 j R_{\text{ref}}^4}{\pi} \frac{\sin(5\alpha_1) - \sin(5\alpha_2) + \sin(5\alpha_3)}{5} \left[\frac{1}{(r+w)^3} - \frac{1}{r^3}\right], \quad (22)$$

$$B_7 = \frac{\mu_0 j R_{\text{ref}}^6}{\pi} \frac{\sin(7\alpha_1) - \sin(7\alpha_2) + \sin(7\alpha_3)}{7} \left[\frac{1}{(r+w)^5} - \frac{1}{r^5}\right].$$

(Note that, conversely, the normalized multipoles are not linear, so it is not permitted to add the contribution of each sector to calculate the final values for the whole coil.) In setting $b_3$ and $b_5$ to zero, we now have a one-parameter family of solutions. Among them, we have three solutions with integer angles that are easy to remember: $(\alpha_1, \alpha_2, \alpha_3) = (24°, 36°, 70°), (36°, 44°, 64°), (48°, 60°, 72°)$. The solution ~$(43.2°, 52.2°, 67.3°)$ sets $b_7$ to zero.

It is now easy to continue: with two wedges, there are five free parameters, and we can set $b_3$–$b_{11}$ to zero. The equations for the angles are as follows:

$$\sin(3\alpha_1) - \sin(3\alpha_2) + \sin(3\alpha_3) - \sin(3\alpha_4) + \sin(3\alpha_5) = 0,$$

$$\sin(5\alpha_1) - \sin(5\alpha_2) + \sin(5\alpha_3) - \sin(5\alpha_4) + \sin(5\alpha_5) = 0,$$

$$\sin(7\alpha_1) - \sin(7\alpha_2) + \sin(7\alpha_3) - \sin(7\alpha_4) + \sin(7\alpha_5) = 0, \quad (23)$$

$$\sin(9\alpha_1) - \sin(9\alpha_2) + \sin(9\alpha_3) - \sin(9\alpha_4) + \sin(9\alpha_5) = 0,$$

$$\sin(11\alpha_1) - \sin(11\alpha_2) + \sin(11\alpha_3) - \sin(11\alpha_4) + \sin(11\alpha_5) = 0.$$

Also, in this case there is one solution with angles 0°–33.3°, 37.1°–53.1°, 63.4°–71.8° (see Fig. 8, left), which is the basis for the RHIC dipole coil layout, where one additional wedge included to recover the correct keystoning of the cables (see Fig. 8, right).

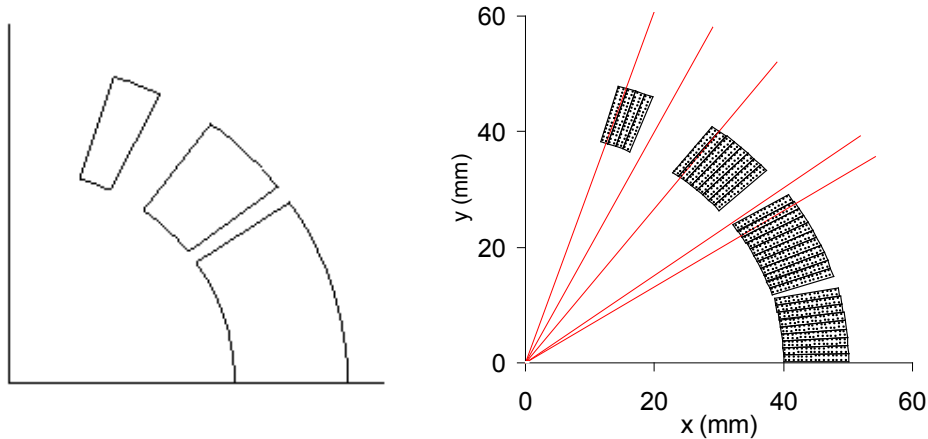

**Fig. 8:** Coil with two wedges which cancels $b_3$ to $b_{11}$ (left), and coil used on the RHIC dipole (right), where the straight lines indicate the angles of the coil.

We now arrive at the goals of this section: estimating the proportionality coefficient between the field, the current density, and the coil width for a few layouts (see Eq. (18) for the 60° sector coil). This allows us to compare how much field we can produce for a given coil width. We define the coefficients $\gamma$ and $\gamma_0$ as

$$B_1 \equiv \gamma j \equiv \gamma_0 w j , \qquad (24)$$

and for a 60° sector coil, see Eq. (18), we calculate

$$\gamma_0 = -\frac{2\mu_0}{\pi}\sin\frac{\pi}{3} = 4\sqrt{3}\times 10^{-7} \approx 6.9\times 10^{-7}\left[\text{T}\cdot\text{m}\cdot\text{A}^{-1}\right]. \qquad (25)$$

The same coefficient can be estimated for layouts with one or more wedges, yielding similar results (see Table 1). It is interesting to note that the addition of one wedge (which allows us to reduce $b_5$ to zero) to layouts with the same angular width (i.e. the same quantity of superconductor) costs only ~5% of the field (from $6.93 \times 10^{-7}$ to $6.63 \times 10^{-7}$; see the first two rows of Table 1). So, in general, the requirements on field quality imply only a small reduction in the field that can be obtained from a generic coil winding.

**Table 1:** Features of six different layouts based on sector coils of increasing complexity

| Layout | No. wedges | Total angular width (degrees) | $\gamma_0$ (T · m /A) | Zero harmonics |
|---|---|---|---|---|
| [0°–60°] | 0 | 60.0 | 6.928E-06 | $b_3$ |
| [0°–48°, 60°–72°] | 1 | 60.0 | 6.625E-06 | $b_3, b_5$ |
| [0°–24°, 36°–60°] | 1 | 48.0 | 5.480E-06 | $b_3, b_5$ |
| [0°–36°, 44°–64°] | 1 | 56.0 | 6.335E-06 | $b_3, b_5$ |
| ~[0°–43.2°, 52.2°–67.3°] | 1 | 58.3 | 6.535E-06 | $b_3, b_5, b_7$ |
| ~[0°–33.3°, 37.1°–53.1°, 63.4°–71.8°] | 2 | 57.7 | 6.411E-06 | $b_3, b_5, b_7$ |

*Example:* The coil of the LHC main dipole [11] is ~30 mm wide and has 350 A·mm$^{-2}$ (inner layer) and 430 A·mm$^{-2}$ (outer layer) as the overall current densities. Applying Eq. (24), using $\gamma_0 = 6.9 \times 10^{-7}$ T·m·A$^{-1}$, one obtains 3.5 T and 4.3 T, thus giving a total of 7.8 T, close to the 8.3 T obtained by an accurate model.

## 2.3 Quadrupoles based on sector coils

The sector layout for a quadrupole is shown in Fig. 9. The arrangement of currents produces a null field in the centre, so $B_1 = 0$, and the second-order harmonic is given by

$$B_2 = -\frac{I\mu_0 R_{\text{ref}}}{2\pi}\text{Re}\left(\frac{1}{z_0^2}\right). \qquad (26)$$

We define the gradient as

$$G \equiv \frac{B_2}{R_{\text{ref}}} = -\frac{I\mu_0}{2\pi}\text{Re}\left(\frac{1}{z_0^2}\right) = -\frac{I\mu_0}{2\pi}\frac{\cos 2\theta}{\rho^2}, \qquad (27)$$

where again we switch to polar coordinates $z = \rho\exp(i\theta)$; on integrating over the sector coil, we obtain

$$I \to j\rho\,d\rho\,d\theta \quad , \quad G = -4\frac{j\mu_0}{2\pi}\int_{-\alpha}^{\alpha}\cos 2\theta\,d\theta\int_{r}^{r+w}\frac{\rho\,d\rho}{\rho^2} = -\frac{2j\mu_0}{\pi}\ln\left(1+\frac{w}{r}\right)\sin 2\alpha. \qquad (28)$$

The gradient is proportional to the current density and to the (natural!) log of one plus the ratio of the coil width and the aperture. This means that

- the gradient is proportional to the current density $j$;
- the gradient depends on the ratio of the coil width $w$ and the aperture $r$;
- for large coil width, the gradient increases with the logarithm of the coil width $\propto \log w$.

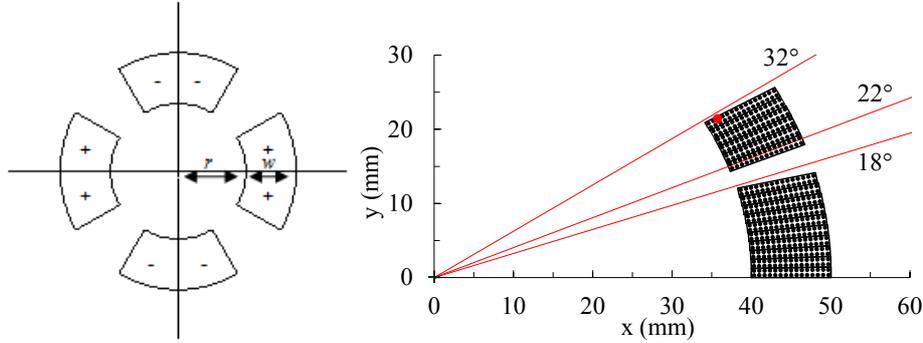

**Fig. 9:** Sector layout for a quadrupole (left), and one-eighth of the RHIC main quadrupole coil, with straight lines indicating the 18°–22°–32° angles (right).

The multipolar expansion is now achieved by factorizing the main component, i.e. $B_2$

$$B(x,y) = 10^{-4} \times B_2 \sum_{n=1}^{\infty}(b_n + ia_n)\left(\frac{x+iy}{R_{\text{ref}}}\right)^{n-1} \quad x,y \in D, \tag{29}$$

and the allowed multipoles are $b_{2n+2}$:

$$B = B_2\left(1 + 10^{-4}\left[b_6 \frac{z^5}{R_{\text{ref}}^5} + b_{10}\frac{z^9}{R_{\text{ref}}^9} + b_{14}\frac{z^{13}}{R_{\text{ref}}^{13}} + \ldots\right]\right). \tag{30}$$

The first piece of good news is that the field quality constraints are easier to calculate for a quadrupole, as there are fewer low-order harmonics to be minimized: keeping the rule of truncation at order 11, one needs to minimize only two harmonics, $b_6$ and $b_{10}$. The second piece of (very) good news is that the equations for the quadrupole field quality are very similar to those for the dipole case, namely any angular solution for a dipole cancelling the first $n$ allowed harmonics will also cancel the first $n$ allowed quadrupolar harmonics, provided that angles are divided by a factor of 2 (see Table 2). Therefore, a 30° sector coil cancels $b_6$, a one-wedge coil [0°–24°, 30°–36°] cancels $b_6$ and $b_{10}$, and so on. So we have already solved the quadrupolar equations for the field quality in the previous subsection (and also for sextupoles, decapoles, etc.)!

**Table 2:** Features of six different layouts based on sector coils of increasing complexity

| Layout | No. wedges | Total angular width (degrees) | $\gamma_0$ (T · m /A) | Zero harmonics |
|---|---|---|---|---|
| [0°–30°] | 0 | 30.0 | 6.928E-06 | $b_6$ |
| [0°–24°, 30°–36°] | 1 | 30.0 | 6.625E-06 | $b_6, b_{10}$ |
| [0°–12°, 18°–30°] | 1 | 24.0 | 5.480E-06 | $b_6, b_{10}$ |
| [0°–18°, 22°–32°] | 1 | 28.0 | 6.335E-06 | $b_6, b_{10}$ |
| ~[0°–21.6°, 26.1°–33.7°] | 1 | 29.2 | 6.535E-06 | $b_6, b_{10}, b_{14}$ |
| ~[0°–16.7°, 18.6°–26.6°, 31.7°-35.9°] | 2 | 28.9 | 6.411E-06 | $b_6, b_{10}, b_{14}$ |

As in the dipole case, we present coil layouts (which are usually selected via numerical codes) that can be understood in terms of the above analytical equations and solutions. We present three examples:

- The one-layer coil of the main RHIC quadrupole [7] that closely follows the [0°–18°, 22°–32°] layout (see Fig. 9, right).

- The two-layer coil of the Tevatron main quadrupole [8], which has an inner layer with a [0°–12°, 18°–30°] layout, and the outer layer is a 30° coil (the outer layer has little impact on higher orders so a wedge to minimize $b_{10}$ is not required); see Fig. 10, left.

- The two-layer coil of the LHC main quadrupole [11], which has an inner layer with a [0°–24°, 30°–36°] layout, and the outer layer is a 30° coil with a small wedge to recover the Roman arch structure; see Fig. 10, right.

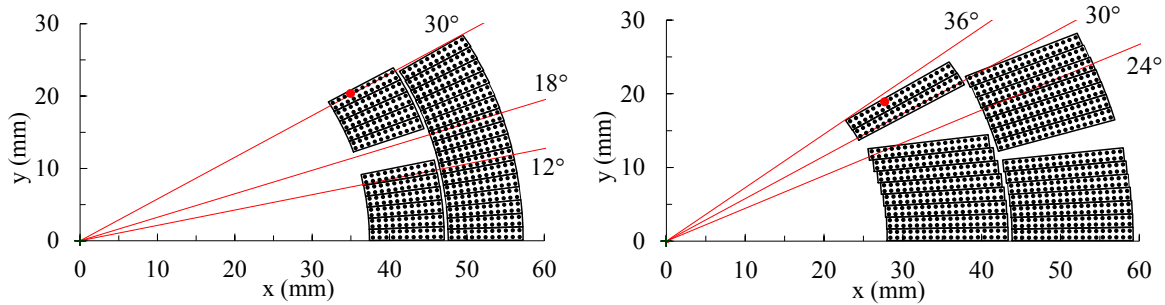

**Fig. 10:** One-eighth of the HERA main quadrupole coil, with straight lines indicating the 12°–18°–30° angles (left), and one-eighth of the LHC main quadrupole coil, with straight lines indicating the 24°–30°–36° angles (right).

We conclude this section as we did the previous one, by estimating the proportionality coefficient between the field, the current density, and the coil width for a few layouts. We define

$$G \equiv \gamma_0 j \ln\left(1 + \frac{w}{r}\right), \tag{31}$$

and we can easily verify that the values for the constant $\gamma_0$ are the same as in the dipole case (provided that angles are divided by 2; see Table 2).

*Example:* The LHC main quadrupole coil has an aperture of 28 mm radius, a width of 31 mm, and 430 A·mm$^{-2}$ overall current density. Applying Eq. (31), using $\gamma_0 = 6.6 \times 10^7$ T·m·A$^{-1}$, we obtain 210 T·m, close to the 220 T·m obtained by an accurate model. Remember, this is a natural logarithm!

## 3 Superconducting magnets

The preceding considerations are valid for any electromagnet based on sector coils. We now consider the case of superconducting sector coils made of Nb–Ti and/or Nb$_3$Sn. We start by defining approximations for the critical surfaces, which are easier to handle than the usual fits and which allow analytical solutions. The magnet is limited not by the field in the centre (dipole) or at the edge of the aperture (quadrupole), but by the peak field in the coil; for this reason, a whole section is dedicated to the estimation of this quantity. Finally, the intersection of the electromagnet loadline $B(j)$ with the superconductor characteristics allows us to derive an analytical expression for the field/gradient as a function of the quantity of coil and the type of superconductor.

## 3.1 Critical surfaces

Superconductors are able to tolerate a given combination of current density and field, in an approximately triangular zone of the $(B, j)$ space, at a fixed temperature [12]. For Nb–Ti, a good approximation is the linear function

$$j_{sc}(B) = s(b - B), \qquad (32)$$

where $b$ and $s$ are two free parameters that fit the critical surface. The fit is accurate over a few tesla (see Fig. 11). Typically, the slope $s \sim 5.0 \times 10^7$ A·m$^{-2}$·T$^{-1}$, i.e. lowering the field by 1 T gains 500 A·mm$^{-2}$. At 1.9 K and 9 T, typical values are ~2000 A·mm$^{-2}$, thus yielding $b \sim 13$ T. At 4.2 K and 6 T, typical values are as well around 2000 A·mm$^{-2}$, thus yielding $b \sim 10$ T, i.e., the whole critical surface at 1.9 K is shifted left by about 3 T A sketch of the critical surface through the Bottura fit [13] and the linear approximation is given in Fig. 11.

For Nb$_3$Sn, the linear approximation is not valid since there is a non-negligible positive curvature [14] (see Fig. 11). A good approximation [6] that provides an analytical solution when intersecting with the magnet loadline is a shifted hyperbola:

$$j_{sc}(B) = s\left(\frac{b}{B} - 1\right). \qquad (33)$$

This is an empirical fit corresponding to a pinning force linear in the field,

$$F(B) = j_{sc}(B) \cdot B = s(b - B), \qquad (34)$$

and works very well over a large range of fields (see Fig. 11). Again, $s$ and $b$ are two free parameters, with the same dimensions as in the previous case (a current density per tesla and a field, respectively). Using $s \sim 3.15 \times 10^9$ A·m$^{-2}$·T$^{-1}$ and $b \sim 22$ T at 4.2 K, we obtain a current density of 2700 A·mm$^{-2}$ at 12 T and of 1450 A·mm$^{-2}$ at 15 T, which are currently close to the best achievable values. Extrapolation at 1.9 K can be achieved with $s \sim 3.3 \times 10^9$ A·m$^{-2}$·T$^{-1}$ and $b \sim 24$ T. As before, these are indicative values, and one can work out the best parameters for each critical surface and range of interest.

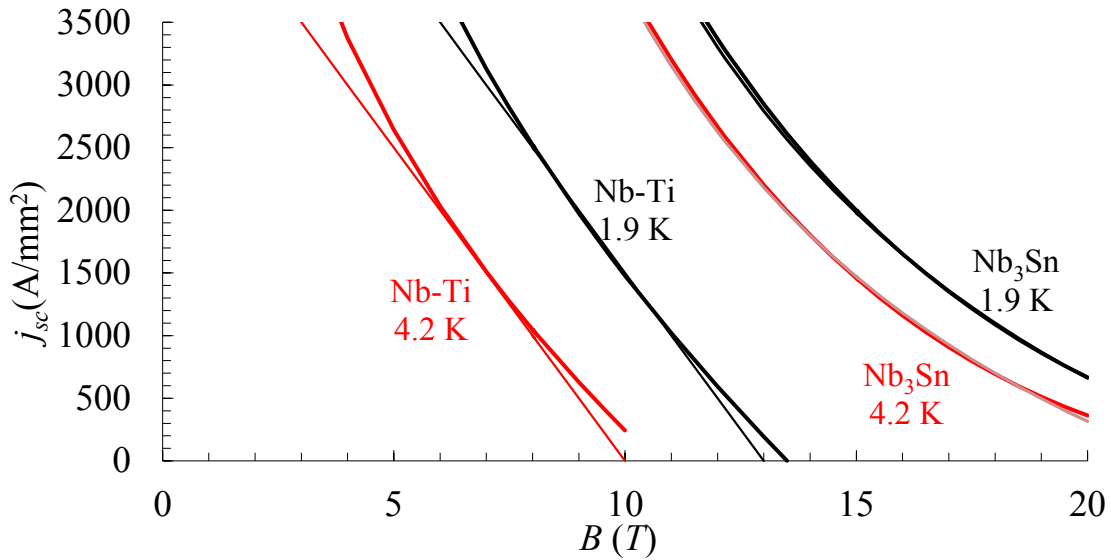

**Fig. 11:** Critical surfaces of Nb–Ti and Nb$_3$Sn; linear fit and hyperbolic fit are shown. (The thin lines in the Nb$_3$Sn case are barely visible as they are very close to the Kramer fit [14].)

The insulated coil contains only a fraction, $\kappa$, of superconductor, with $\kappa$ usually in the range 0.25–0.40. So the equation for the overall current density (including stabilizer, voids, and insulation) is

$$\boxed{j(B) = \kappa s(b - B)} \quad \text{for Nb–Ti} \tag{35}$$

and

$$\boxed{j(B) = \kappa s\left(\frac{b}{B} - 1\right)} \quad \text{for Nb}_3\text{Sn}. \tag{36}$$

### 3.2 Peak field in the coil

The superconducting magnet is limited by the combination of current density and magnetic field in the coil. Because of this, an estimate of the peak field in the coil is required. For a dipole, we define $\lambda$ as the ratio between the peak field $B_p$ and the central field $B_1$:

$$B_p \equiv \lambda B_1. \tag{37}$$

One can show that $\lambda$ is a function of the ratio of the coil aperture and the coil width, i.e. $w/r$. A rough estimate based on the fit of many superconducting coils of accelerator magnets [6] is given by

$$\lambda = 1 + a\frac{r}{w}, \tag{38}$$

with $a \sim 0.04$ (see Fig. 12). This means that for $r \sim w$ (as in the LHC main dipole), the field in the coil is about 4% larger than the central field. The estimate becomes rather pessimistic for $w/r < 0.5$, where it has been shown that an improved result may be achieved with careful coil optimization. For example, for the FAIR dipole with $r = 75$ mm and $w = 15$ mm, the proposed coil layout has $\lambda = 1.13$ instead of the 1.20 expected from our fit.

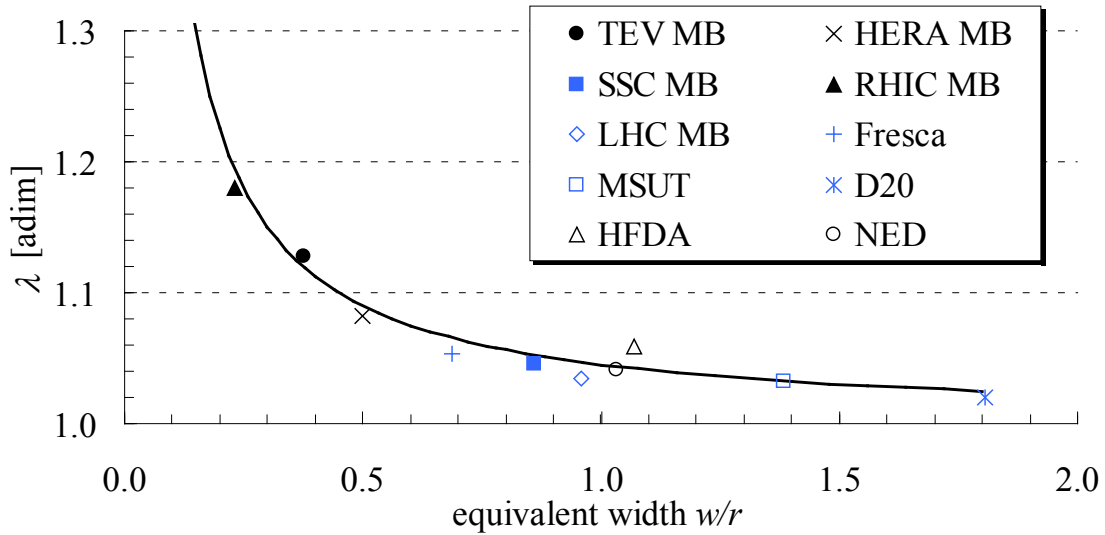

**Fig. 12:** Ratio between peak field and central field in several accelerator magnet dipoles, and hyperbolic fit with $a = 0.04$.

For a perfect quadrupole, the peak field on the border of the aperture is $rG$, so one can define a similar dimensionless ratio $\lambda$ as follows:

$$B_p \equiv \lambda G r . \tag{39}$$

In addition, (i) $\lambda$ is a function of $w/r$, and (ii) $\lambda$ increases for $w/r \ll 1$; however, there is a new feature: for larger and larger coil widths, one does not reach the ideal case $\lambda = 1$, rather there is another divergence. The following fit has been proposed [5]:

$$\lambda = 1 + a_1 \frac{r}{w} + a_{-1} \frac{w}{r}, \qquad (40)$$

with $a_1 = 0.04$ and $a_{-1} = 0.11$, based on the data of several coils (see Fig. 13). These fits have to be considered as approximated analytical expressions, allowing for the fact that $\lambda \neq 1$.

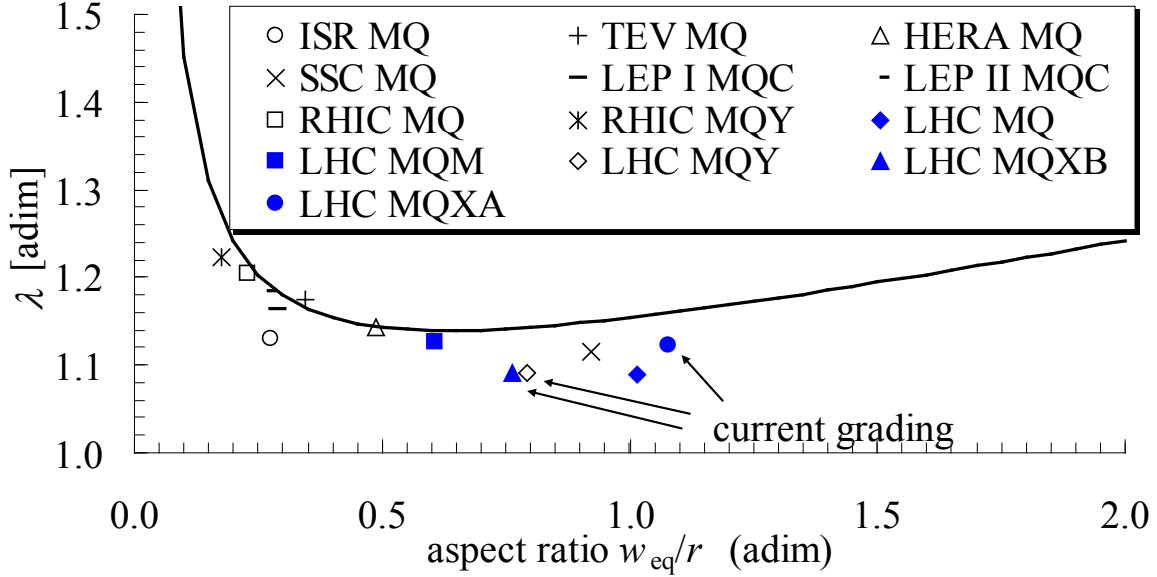

**Fig. 13:** Ratio of peak field and gradient times aperture in several accelerator magnet quadrupoles, and fit of Eq. (40).

### 3.3 Field versus coil width in superconducting dipoles

One of the central choices in magnet design is the selection of the width of the coil, i.e. how much superconductor to use in the magnet. For a generic electromagnet, this choice affects the current density and the operational field through Eq. (24). In a superconducting magnet, current density and field are not independent variables due to the constraints imposed by the critical surface (see Fig. 11). A magnet with small coil width will produce little field for a given current density, so the loadline will be steep. This magnet will reach the critical surface in a 'low field' and large current density region (see Fig. 14, left). On the other hand, a magnet with large coil width will produce a high field for the same current density, and the loadline will have a very gentle slope. This magnet will reach the critical surface at high field and low current density.

We now make these arguments more quantitative by computing the operational limits for the Nb-Ti case. On intersecting the loadline

$$B_p = \lambda \gamma_0 j w \qquad (41)$$

with the linearized critical surface (32) and solving for $j$, we obtain

$$j = \kappa s \left[ b - B_p(j) \right] = \kappa s \left[ b - \lambda \gamma_0 j w \right] = \kappa s b - \kappa s \lambda \gamma_0 j w \quad \rightarrow \quad j(1 + \kappa s \lambda \gamma_0 w) = \kappa s b. \quad (42)$$

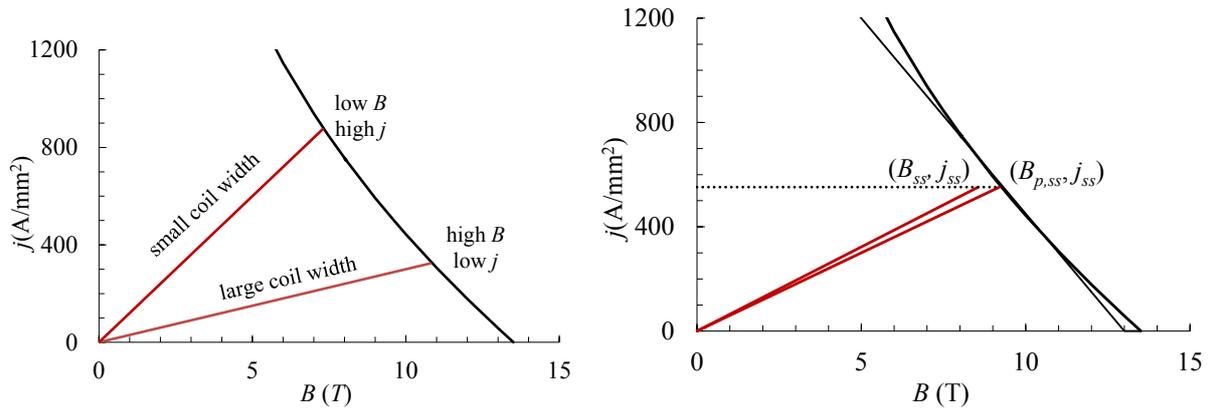

**Fig. 14:** Magnet loadlines (red line) and Nb–Ti critical surface, for large and small coil widths (left), and definition of short sample current density, central field and peak field (right).

The so-called short sample condition, i.e. the limit of the magnet given by the conductor performance (see Fig. 14, right), is obtained:

$$B_{ss} = \frac{\kappa s \gamma_0 w}{1 + \kappa s \lambda \gamma_0 w} b \quad\quad B_{p,ss} = \frac{\kappa s \lambda \gamma_0 w}{1 + \kappa s \lambda \gamma_0 w} b \quad\quad j_{ss} = \frac{\kappa s}{1 + \kappa s \lambda \gamma_0 w} b. \quad (43)$$

The main feature of these expressions is that for $w \to \infty$ the central field reaches the asymptotic limit $b$ as $w/(1+w)$. This sets a practical limit on the maximum field achievable with Nb–Ti of ~10 T, not at $b \sim 13$ T. The remaining 2–3 T can be achieved only with very large coils, which become more and more expensive. The LHC dipole coil, with its coil of 30 mm width, is therefore at the limit for Nb–Ti. To gain 1 T in the short sample field, we would need to increase the coil width from 30 to 50 mm (see Fig. 15). Equations (43) also give the dependence on the filling ratio, on the critical surface properties, and the second-order dependence on the magnet aperture.

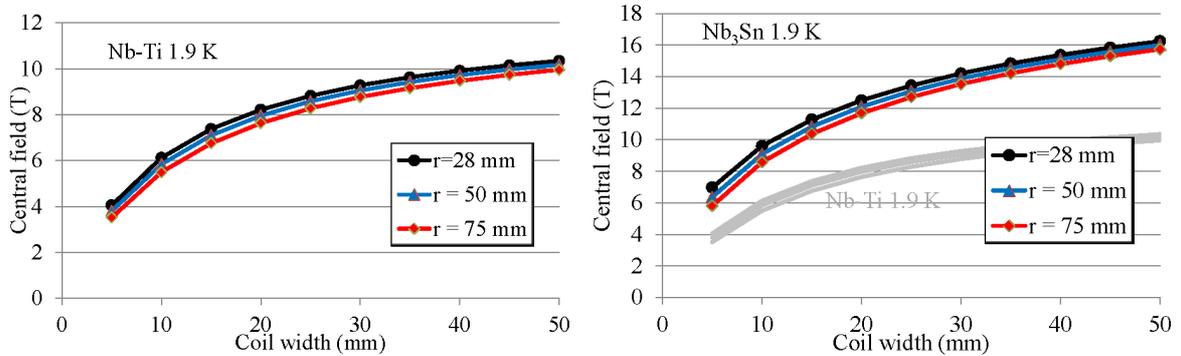

**Fig. 15:** Short sample field versus coil width and aperture for Nb–Ti (left) and Nb$_3$Sn (right); filling factor $\kappa = 0.3$.

*Example:* Before switching to the Nb$_3$Sn case, we apply Eqs. (43) to the case of the LHC main dipole [dip]. We have $w = 31$ mm, $r = 28$ mm, and therefore

$$\lambda = 1 + a \frac{r}{w} = 1 + 0.04 \frac{28}{31} = 1.036, \quad (44)$$

from Eq. (38). The filling factor equals 0.28 and 0.33 in the inner and outer layers, respectively, so we choose an average value $\kappa = 0.3$. Using $s = 5.0 \times 10^8$ A·m$^{-2}$ and $b = 13$ T, we have

$$B_{ss} = \frac{\kappa s \gamma_0 w}{1 + \kappa s \lambda \gamma_0 w} b = \frac{0.3 \times 5.0 \times 10^8 \times 6.6 \times 10^{-7} \times 0.031}{1 + 0.3 \times 5.0 \times 10^8 \times 1.036 \times 6.6 \times 10^{-7} \times 0.031} 13 = 9.6 \text{ T}, \quad (45)$$

which is very close to the real value of 9.7 T, which also includes the iron effect (see next section), and

$$j_{ss} = \frac{B_{ss}}{\gamma_0 w} = \frac{9.6}{6.6 \times 10^{-7} \times 0.031} = 4.7 \times 10^8 \text{ A} \cdot \text{m}^{-2}, \quad (46)$$

i.e. 470 A·mm$^{-2}$. (Remember to keep the SI units, even though they are not practical for our case of coil width and aperture in metres, current density in ampere per metre squared.)

For Nb$_3$Sn, Eq. (35) for the critical surface must be replaced by Eq. (36). The hyperbolic fit gives a second-order equation that can be solved explicitly:

$$\boxed{B_{ss} = \frac{\kappa s \gamma_0 w}{2}\left(\sqrt{\frac{4b}{\kappa s \lambda \gamma_0 w} + 1} - 1\right)}, \quad j_{ss} = \frac{B_{ss}}{\gamma_0 w}, \quad B_{p,ss} = \lambda B_{ss}. \quad (47)$$

For $w \to \infty$ in this case, we reach the field limit $b$; saturation is slower w.r.t. Nb-Ti (see Fig. 15) due to the positive curvature of the critical surface.

*Example:* We consider the case of an 11 T dipole, with a coil geometry very similar to the LHC main dipole: $r = 30$ mm, $w = 31$ mm. The peak to bore field ratio is given by

$$\lambda = 1 + a\frac{r}{w} = 1 + 0.04\frac{30}{31} = 1.039. \quad (48)$$

The short sample field is around 14 T:
$$\gamma_0 w = 6.6 \times 10^{-7} \times 0.031 = 2.05 \times 10^{-8};$$

$$B_{ss} = \frac{0.3 \times 3.3 \times 10^9 \times 2.05 \times 10^{-8}}{2}\left(\sqrt{\frac{4 \times 24}{0.3 \times 3.3 \times 10^9 \times 1.039 \times 2.05 \times 10^{-8}} + 1} - 1\right) = 13.8 \text{ T}, \quad (49)$$

which becomes 11 T when the 20% margin is imposed. The short sample current is 670 A·mm$^{-2}$:

$$j_{ss} = \frac{B_{ss}}{\gamma_0 w} = \frac{13.8}{6.6 \times 10^{-7} \times 0.031} = 6.7 \times 10^8 \text{ A} \cdot \text{m}^{-2}. \quad (50)$$

### 3.4 Field versus coil width in superconducting quadrupoles

The same approach can be used to work out the quadrupole case [6]. For the Nb–Ti,

$$\boxed{G_{ss} = \frac{\kappa s \gamma_0 \ln(1 + w/r)}{1 + \kappa r s \lambda \gamma_0 \ln(1 + w/r)} b} \quad \boxed{j_{ss} = \frac{G_{ss}}{\gamma_0 \ln(1 + w/r)}} \quad \boxed{B_{p,ss} = \lambda r G_{ss}}, \quad (51)$$

where $\lambda$ is given by Eq. (40). A plot giving the short sample gradient versus the coil width and the aperture is shown in Fig. 16. The Nb$_3$Sn technology gives ~50% more gradient for the same aperture w.r.t. Nb-Ti.

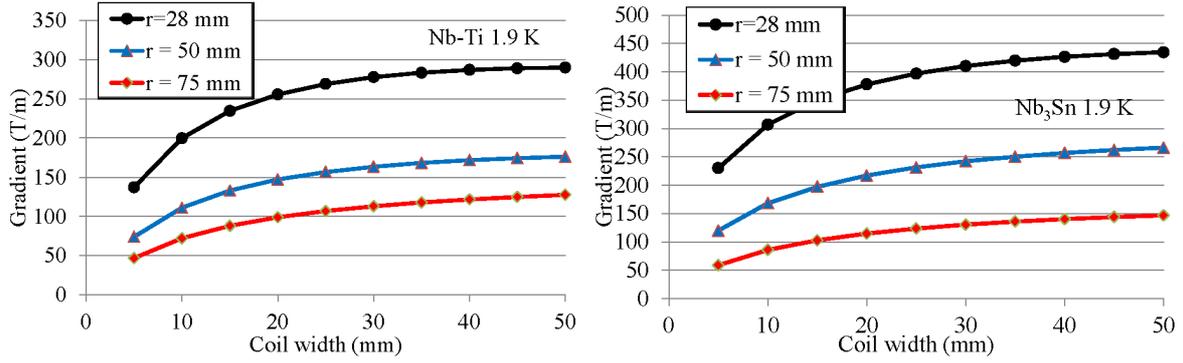

**Fig. 16:** Short sample gradient versus coil width and aperture for Nb–Ti and $Nb_3Sn$; filling factor $\kappa = 0.3$

*Example:* We apply this equation to the case of the large-aperture Nb-Ti quadrupole developed for the LHC upgrade. It has aperture $r = 60$ mm and coil width $w = 31$ mm (two layers of the LHC main dipole cable), and therefore the peak field to bore field ratio is given by

$$\lambda = 1 + a_1 \frac{r}{w} + a_{-1} \frac{w}{r} = 1 + 0.04 \frac{60}{31} + 0.11 \frac{31}{60} = 1.13. \tag{52}$$

The log factor is

$$\ln\left(1 + \frac{w}{r}\right) = \ln\left(1 + \frac{31}{60}\right) = 0.42, \tag{53}$$

and therefore one has

$$G_{ss} = \frac{0.3 \times 5.0 \times 10^8 \times 6.6 \times 10^{-7} \times 0.42}{1 + 0.3 \times 0.06 \times 5.0 \times 10^8 \times 1.13 \times 6.6 \times 10^{-7} \times 0.42} 123 = 141\, \text{T·m}^{-1}, \tag{54}$$

quite close to the real value of 149 T·m$^{-1}$. The current density and peak field are

$$j_{ss} = \frac{141}{6.6 \times 10^{-7} \times 0.42} = 5.1 \times 10^8\, \text{A·m}^{-2} \quad \text{and} \quad B_{p,ss} = 1.13 \times 0.06 \times 141 = 9.6\, \text{T}. \tag{55}$$

For the $Nb_3Sn$ case, the equations become more cumbersome, but are still manageable in a spreadsheet:

$$\boxed{G_{ss} = \frac{\kappa s \gamma_0 \ln(1 + w/r)}{2} \left( \sqrt{\frac{4b}{\kappa r s \lambda \gamma_0 \ln(1 + w/r)} + 1} - 1 \right)}, \tag{56}$$

and the same expressions hold for the short sample peak field and current density as in Eq. (51).

*Example:* We consider the case of the $Nb_3Sn$ quadrupole for the LHC upgrade QXF, having $r = 75$ mm and $w = 36$ mm, and therefore

$$\lambda = 1 + a_1 \frac{r}{w} + a_{-1} \frac{w}{r} = 1 + 0.04 \frac{75}{36} + 0.11 \frac{36}{75} = 1.14. \tag{57}$$

The log factor is

$$\ln\left(1 + \frac{w}{r}\right) = \ln\left(1 + \frac{36}{75}\right) = 0.39, \tag{58}$$

and therefore we have

$$\kappa s \gamma_0 \ln(1+w/r) = 0.32 \times 3.3 \times 10^9 \times 6.6 \times 10^{-7} \times 0.39 = 274 \text{ T} \cdot \text{m}^{-1}, \tag{59}$$

$$G_{ss} = \frac{274}{2}\left(\sqrt{\frac{4 \times 24}{274 \times 1.13 \times 0.075}+1}-1\right) = 173 \text{ T} \cdot \text{m}^{-1}, \tag{60}$$

very close to the real value of 169 T·m$^{-1}$. The current density and peak field are

$$j_{ss} = \frac{173}{6.6 \times 10^{-7} \times 0.39} = 6.7 \times 10^8 \text{ A} \cdot \text{m}^{-2} \quad \text{and} \quad B_{p,ss} = 1.14 \times 0.075 \times 173 = 14.7 \text{ T}. \tag{61}$$

## 4 The role of iron

Iron has several functions: (i) to keep the magnetic flux within the magnet, (ii) to prevent fringe fields in the tunnel that may endanger the electronics, and (iii) to reduce the current density in the magnet via the creation of a virtual coil. Since the iron can retain a maximum of $B_s \sim$ 1.5–2 T, for a dipole the iron thickness $t$ needed for shielding is

$$\boxed{t \sim \frac{rB_1}{B_s}}. \tag{62}$$

For example, in the LHC dipole we have $r$ = 28 mm and $B_1$ = 8.3 T, yielding $t \sim$ 100 mm. For a quadrupole, the shielding condition is given by

$$\boxed{t \sim \frac{r^2 G}{2B_s}}. \tag{63}$$

So, for the LHC main quadrupole, $G$ = 220 T·m$^{-1}$, $r$ = 28 mm, and $t \sim$ 40 mm.

Inside the aperture, the iron acts as a virtual second-layer coil (see Fig. 17). According to the current image method (see, e.g., [1], p. 53), a thick ring of iron at a distance $R_I$ from the centre of the aperture produces an image current of a current line $\rho$, which is at a distance $\rho'$ and carries current $I'$, where

$$\rho' = \frac{R_I^2}{\rho}, \qquad I' = \frac{\mu - 1}{\mu + 1} I \sim I. \tag{64}$$

This approximation is valid for non-saturated iron. Above 2 T, iron becomes saturated, and the effect of the virtual coil decreases. In this case, a finite element code [15, 16] should be used to construct the calculations. Geometries that do not satisfy the circular symmetry in the non-saturated case can be treated as a first approximation with an radius of an iron ring. For non-saturated iron, integrating over a sector coil of width $w$ in an aperture $r$ leads to a virtual coil of inner radius $R_1$ and outer radius $R_2$ (see Fig. 17, right), giving an additional field

$$\Delta B_1 = \gamma_0 J(R_2 - R_1), \tag{65}$$

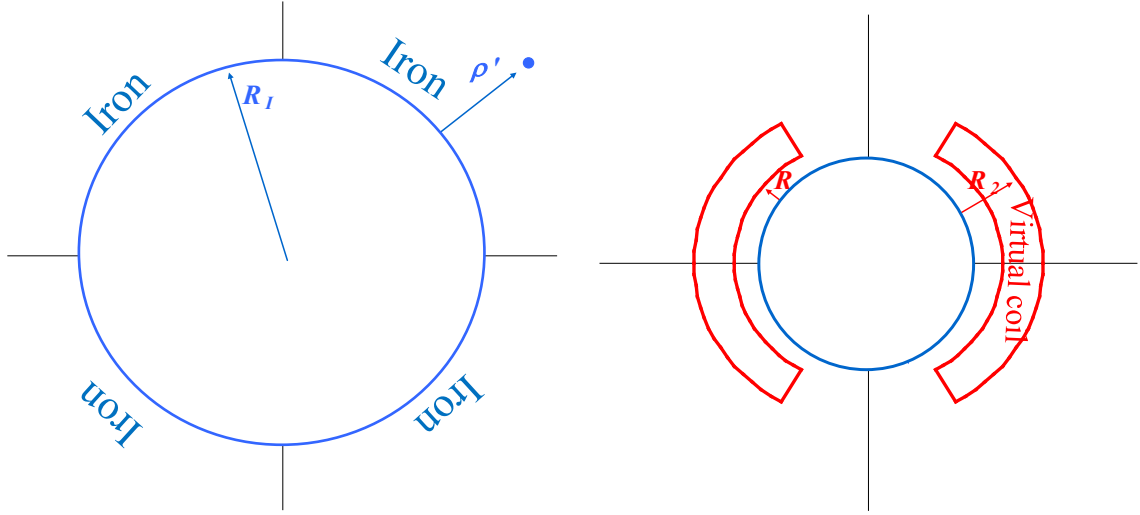

**Fig. 17:** Impact of iron: current image produced by iron at $R_I$ (left), and sector coil with virtual coil provided by iron (right).

Note that the current densities of the virtual coil $J$ and of the coil $j$ are not the same, since the total current is the same but the area of the virtual coil is much larger:

$$j\left[(r+w)^2 - r^2\right] = J(R_2^2 - R_1^2) \quad \rightarrow \quad J(R_2 - R_1) = j\frac{(r+w)^2 - r^2}{R_2 + R_1}; \tag{66}$$

therefore, substituting Eq. (66) into Eq. (65) we obtain,

$$\Delta B_1 = \gamma_0 j \frac{(r+w)^2 - r^2}{R_2 + R_1}, \tag{67}$$

and, since $R_2 = R_I^2/r$ and $R_1 = R_I^2/(r+w)$,

$$\Delta B_1 = \gamma_0 j \frac{(r+w)^2 - r^2}{R_2 + R_1} = \frac{\gamma_0 j}{R_I^2} \frac{(r+w)^2 - r^2}{\frac{1}{r} + \frac{1}{r+w}} = \frac{\gamma_0 j}{R_I^2} wr(r+w). \tag{68}$$

Finally, after all this algebra, we obtain a beautiful, simple expression: the iron simply increases the field of a fraction $f$, which is a function of the aperture, of the coil width, and of the inner radius of the iron:

$$\boxed{f \equiv \frac{\Delta B_1}{B_1} = \frac{r(r+w)}{R_I^2}}. \tag{69}$$

This equation is a generalization of the estimate for a thin coil [1], p. 54, to a thick coil. In the most favourable case $w \ll r$ and $R_I = r + w$ (thin coil surrounded by iron without spacers or collars), the iron can double the main field. Typically, one needs some space between the coil and the iron (collars or spacers 10 to 50 mm thick, see Fig. 18), and typical gains are in the range 10–50%. The largest gain is not only for small collars, but also for thin coils, as in the RHIC case (see Fig. 19). In the LHC case, the gain is ~17%.

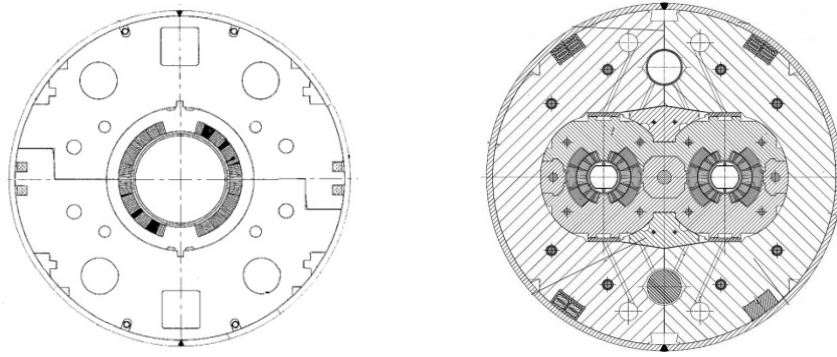

**Fig. 18:** Cross-section of the RHIC dipole [7] (left), with thin spacers, and of the LHC dipole [11] (right), with thick collars.

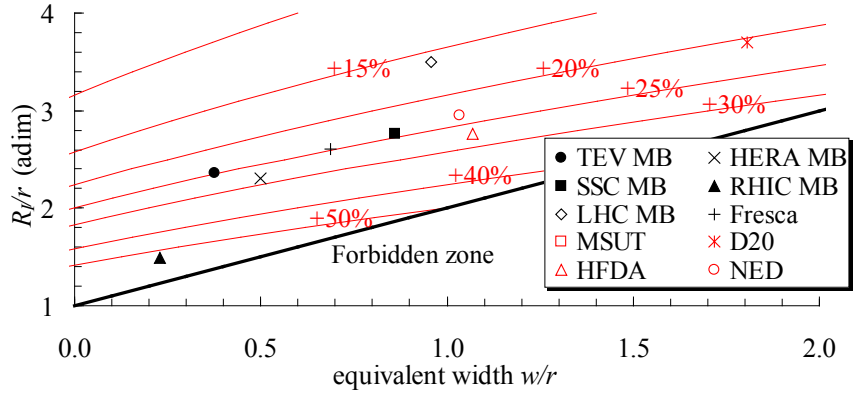

**Fig. 19:** Impact of iron: increase in main field vs. iron radius and coil width

Iron not only increases the field for a given current density, but also lowers the magnet loadline, thereby increasing the short sample field: the effect of including the iron is to increase the coefficient $\gamma_0$ of the fraction $f$ (see Eq. (69)). For example, in a Nb-Ti dipole we have

$$B_{ss} = \frac{\kappa s \gamma_0 (1+f) w}{1 + \kappa s \lambda \gamma_0 (1+f) w} b, \qquad f = \frac{r(r+w)}{R_I^2}. \tag{70}$$

*Example:* We can recompute the LHC dipole case adding the 17% to obtain

$$B_{ss} = \frac{0.3 \times 5.0 \times 10^8 \times 6.6 \times 10^{-7} \times 1.17 \times 0.031}{1 + 0.3 \times 5.0 \times 10^8 \times 1.036 \times 6.6 \times 10^{-7} \times 1.17 \times 0.031} 13 = 9.9 \text{ T}, \tag{71}$$

i.e. 3% more of that estimated without iron. Indeed, there is a beneficial decrease of the short sample current, which decreases from 470 A·mm$^{-2}$ (see Eq. (46)) to

$$j_{ss} = \frac{B_{ss}}{\gamma_0 w} = \frac{9.9}{6.6 \times 10^{-7} \times 1.17 \times 0.031} = 4.1 \times 10^8 \text{ A} \cdot \text{m}^{-2}; \tag{72}$$

this reduces stresses and provides a greater margin for magnet protection.

*Example:* It is interesting to consider the extreme case of the RHIC dipole ($r = 40$ mm, $w = 10$ mm, filling factor $\kappa = 0.22$, operational temperature 4.2 K); without iron, we would have a short sample field of

$$B_{ss} = \frac{0.22 \times 5.0 \times 10^8 \times 6.6 \times 10^{-7} \times 0.010}{1 + 0.22 \times 5.0 \times 10^8 \times 1.16 \times 6.6 \times 10^{-7} \times 0.010} 9 = 3.6 \text{ T}, \tag{73}$$

and with the iron we obtain

$$B_{ss} = \frac{0.22 \times 5.0 \times 10^8 \times 6.6 \times 10^{-7} \times 1.56 \times 0.010}{1 + 0.22 \times 5.0 \times 10^8 \times 1.16 \times 6.6 \times 10^{-7} \times 1.56 \times 0.010} 9 = 4.4 \text{ T}, \qquad (74)$$

i.e. a considerable gain of 20% additional short sample field. The short sample current without iron is

$$j_{ss} = \frac{B_{ss}}{\gamma_0 w} = \frac{4.4}{6.6 \times 10^{-7} \times 0.010} = 5.4 \times 10^8 \text{ A} \cdot \text{m}^{-2}, \qquad (75)$$

and with the iron it decreases by 20%, from 540 A·mm$^{-2}$ to 430 A·mm$^{-2}$:

$$j_{ss} = \frac{B_{ss}}{\gamma_0 w} = \frac{4.4}{6.6 \times 10^{-7} \times 1.56 \times 0.010} = 4.3 \times 10^8 \text{ A} \cdot \text{m}^{-2}. \qquad (76)$$

## 5  Other layouts

The sector coil is the workhorse of accelerator magnets. All superconducting magnets installed in an accelerator are based on the cos-theta layout, which is achieved using sector coils. Alternative layouts [17, 18] have been proposed and implemented in short models, with mixed results. Here we briefly discuss the block coil, and how to extend the previous equations to the case of a layout not based on cos-theta. The winding of a block coil is similar to that of a solenoid: cables are perpendicular to the horizontal midplane, rather than being parallel as in the cos-theta layout. Moreover, cables are not keystoned, and they form rectangular blocks arranged in layers. If a layer is above the beam tube, it can be wound as a flat racetrack coil, but if it is close to the midplane the end must be opened to make space for the beam tube (flared ends). An example is the beautifully simple layout of HD2 (see Fig. 20).

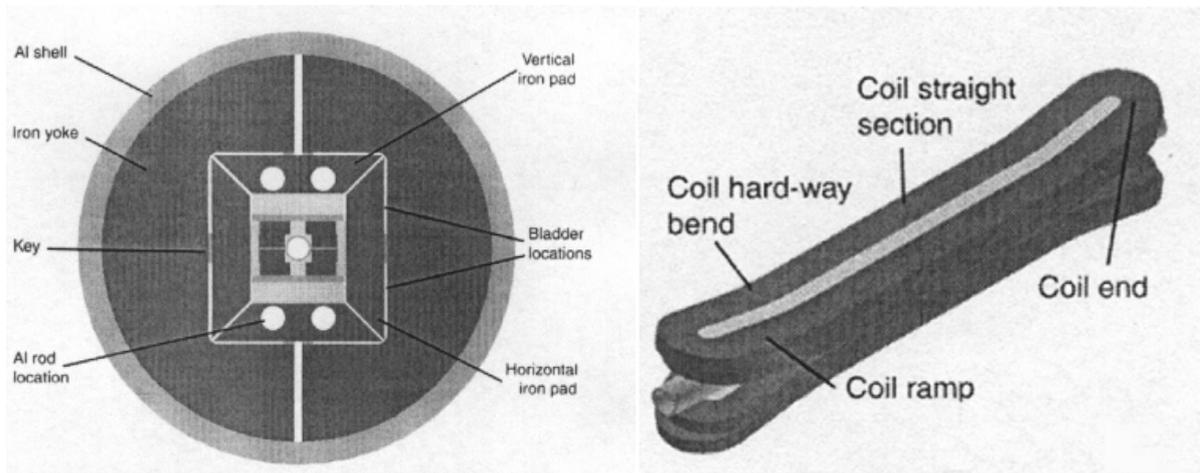

**Fig. 20:** HD2, a dipole based on the block concept

At present, there is no consensus on the feasibility and advantages of the block layout. We make the following remarks, although we are aware that some of them are not universally shared.

- The main feature is that the cos-theta layout is self-supporting, with no need for an internal structure. On the other hand, the block needs an inner tube to support the forces acting on the coil, reducing the aperture available to the beam.

- The block layout is not viable for a small coil to aperture ratio $w/r$, as in the RHIC magnet. If the coil is thin, it does not make sense to arrange the conductors in a block geometry. On the other hand, for large $w/r \gg 1$ (~2 in HD2), the block coil becomes an interesting option.

- Both layouts are equivalent from the point of view of field quality optimization. Multipoles can be cancelled with a few free parameters, i.e. by adding wedges. Note that the HD2 case does not need wedges because the ratio $w/r$ is large; see Eq. (19).

- The block layout advantage of a simple coil geometry, squared and without wedges, is apparent: the winding of cos-theta coils with wedges is a well-mastered technology, including the ends. On the other hand, for the block layout the mechanical support of the coil ends and the transition in the flared region have not yet been mastered. To be fair, one also has to take into account the fact that we are comparing a few block short models with several thousands of cos-theta magnets.

- The forces accumulate on the midplane in a cos-theta geometry, whereas in a block coil they occur mainly in the horizontal direction.

- With respect to the cos-theta layout, the block has a somewhat larger value of $\lambda$ and a lower $\gamma_0$, so the design is less effective [19]. This difference vanishes for very large coils.

- The block layout is very flexible, allowing the number of turns in the outer part to be adjusted, whereas for a cos-theta layout one has to add a whole layer (or change the cable width). Since the level curves of the field are mostly parallel to the vertical direction, in the block layout it is easier to have material grading.

One can use the equations derived in this paper for block layouts (or for any layout) by replacing the coil width $w$ by an equivalent coil width $w_e$, defined as the width of the 60° sector coil which has the same quantity of superconductor of the 'exotic' design, and the same aperture $r$. Let $A$ be the cross-sectional surface area of the coil; we equate it to the surface area of a sector coil as follows:

$$A \equiv \frac{2\pi}{3}\left[(r+w_e)^2 - r^2\right]. \tag{77}$$

From this definition, we obtain

$$\frac{3A}{2\pi} = r^2\left[\left(1+\frac{w_e}{r}\right)^2 - 1\right], \quad \text{i.e.} \quad \boxed{w_e = \left(\sqrt{1+\frac{3A}{2\pi r^2}} - 1\right)r}. \tag{78}$$

*Example:* HD2 has a 20 mm aperture, and makes use of rectangular insulated cable of 22.2 mm width and 1.62 mm thickness, i.e. 36 mm². There are 24 and 30 turns in each layer, thus giving a total coil cross-sectional surface area of 7800 mm². This gives

$$w_e = \left(\sqrt{1+\frac{3\times 7800}{2\pi \times 20 \times 20}} - 1\right)\times 20 = 44 \text{ mm}. \tag{79}$$

## 6 Putting it all together

The starting point for magnet design is the aperture. This is usually a factor that we accept as an input to our calculations, i.e. it is a requirement coming from accelerator physicists. The main choices are:

- technology and operational temperature, which fix a maximum attainable field;
- bore field or gradient;
- current density;
- width of the coil, i.e. how much superconductor to use;
- fraction of stabilizer, which dilutes the superconductor.

These five quantities are related via equations that can be implemented in a spreadsheet, namely (as discussed in Section 3), Eqs. (43) and (47) for dipoles, and Eqs. (51) and (56) for quadrupoles, for the Nb–Ti and Nb$_3$Sn cases, respectively. Figures 14 (dipoles) and 15 (quadrupoles) give these coupled dependences in the parameter space for accelerator magnets, and can be a useful instrument for a quick estimate of what can be done, what is better not to try, and what is impossible. Out of the above-mentioned quantities, the current density is probably the most relevant parameter in magnet design: it affects many different aspects, such as protection, mechanical structure and stresses, and stability. Large current densities provide more compact magnets, but as can be observed from examining the history of accelerator magnets, most cases range between 300 and 500 A·mm$^{-2}$ (see Figs. 21 and 22).

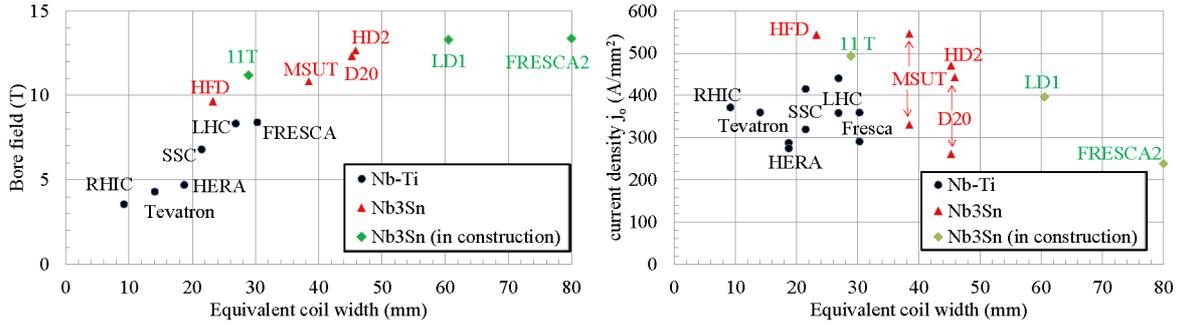

**Fig. 21:** Bore field and current density (both at 20% margin) vs. coil width in dipoles

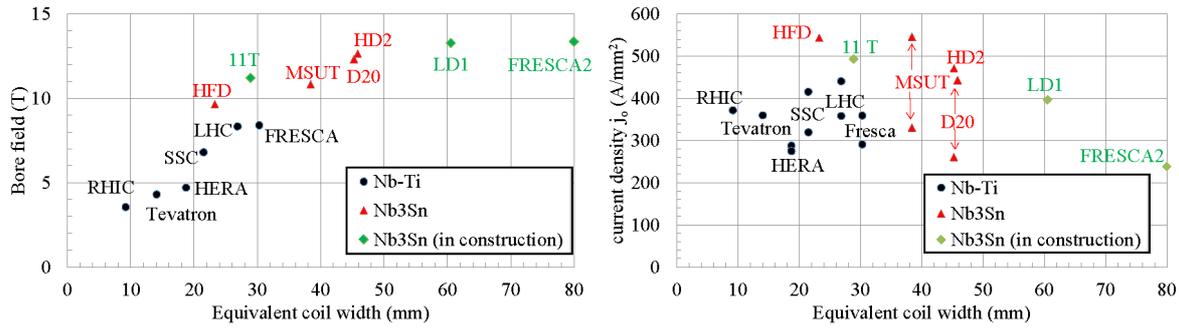

**Fig. 22:** Peak field and current density (both at 20% margin) vs. coil width in quadrupoles

Given the choice of field (peak field for quadrupoles) and coil width, one has to satisfy field quality conditions using the material presented in Section 2. An adequate number of layers and wedges are needed to tune the field harmonics; on the other hand, the unnecessary complexity of too many layers and wedges should be avoided. The main principles of field quality optimization for dipoles and quadrupoles, pointing out the main features, have been discussed. When taking into account that we have a finite size cable and imperfect keystoning, plus other effects not discussed here, such as persistent current and saturation, one has to implement a second fine tuning of the field quality. This is achieved via a code, sometimes through optimization algorithms, or by hand. After construction, a third iteration is usually necessary to apply another fine tuning of the field quality. The capability to go through successive approximations is a fundamental approach of magnet design.

## Acknowledgements

We wish to thank D. Tommasini, M. Juchno, and S. Izquerdo Bermudez for their critical reading of the manuscript and valuable comments. These notes are based on material prepared for the USPAS lectures, in collaboration with H. Felice, P. Ferracin, and S. Prestemon.